\documentstyle[12pt,aps,psfig]{revtex}
\begin{document}
\draft
\title{\bf{ Deadlocks and waiting times in traffic jam } }
\author{Sutapa Mukherji\cite{eml1} and Somendra M.
Bhattacharjee\cite{eml2}} 
\address{Institute of Physics, Bhubaneswar 751 005, India}
\maketitle
\widetext
\begin{abstract}
In a city of right moving and upmoving cars with hardcore constraint,
traffic jam occurs in the form of bands.  We show how the bands are
destroyed by a small number of strictly left moving cars yielding a
deadlock phase with a rough edge of left cars.  We also show that the
probability of waiting time at a signal for a particular tagged car
has a power law dependence on time, indicating the absence of any
characteristic time scale for an emergent traffic jam.  The exponent
is same for both the band and the deadlock cases. The significances of
these results are discussed.
\end{abstract}

It is a common knowledge that vehicular traffic in a city can form a
traffic jam for high enough density of cars, even when the traffic
rules are obeyed.  The jammed phase involves cars segregating and
blocking one another over a large length scale and time, reminiscent
of cooperative phenomena.  Such jams are rather ubiquitous in various
situations from traffic flow to data transmission on a network, phase
separation in granular materials, etc.  In order to identify and
characterize the basic features of traffic jams, simple cellular
automata (CA) inspired models
\cite{biham,naga,nagb,cuesta,martinez,molera,nagela,nagelb,bando,kert1,kert2}
have gained importance over the last few years.  Indeed, such CA
models indicate that a traffic jam can be viewed as a dynamic
transition.  Apart from the practical relevance, traffic jam models
constitute a special class in the general framework of driven
systems. The obvious interest in such studies is in the nonequilibrium
steady states, effects of broken ergodicity, nonequilibrium phase
transitions etc\cite{ziadl}.
  
The CA approach has so far been of two types: (1) one dimensional for
highway traffic \cite{nagela,nagelb,bando,kert1} and (2) two
dimensional for city traffic
\cite{biham,naga,nagb,cuesta,martinez,kert2}, with distinct inputs for
the two types.  As yet, the role of dimensionality in traffic jam is
not clear and, from the models, one might suspect that the traffic
jams in 1-d are rather different from that in two dimensions.  It is
also tempting to propose that the three or higher dimensional models
might be relevant to communication network, computer architecture etc.

In this paper, we consider only the two dimensional city traffic case.
In the CA approach \cite{biham,naga,nagb,cuesta,martinez}, a city is
taken as a square lattice with, say two types of cars, one set moving
right and the other going up. A hard core constraint restricts
occupation of two cars at the same site, preventing accidents.  There
is also a signal at each site (crossing) regulating traffic
synchronously so that cars can move only right in the odd steps and
only up in the even steps. See Fig. 1. A stochasticity is introduced
by allowing a right (up) car, signal permitting, to go up (right) with
probability $\gamma$.  One starts with a random homogeneous initial
condition and let the system evolve to its steady state according to
the stochastic rules\cite{comm}.  Note also that the rules ensure that
the jams are not formed due to stray incidents like accidents or
peculiar signals.

The deterministic model ($\gamma =0$) was shown to have a traffic jam
phase\cite{biham,naga,nagb} with no particular structure, while in the
stochastic case ($\gamma \neq 0$), the jammed state involves well
defined bands with cars of two types phase segregated, blocking one
another \cite{molera,comm1} provided the density, $n$, of cars is
greater than a critical value $n_c(\gamma)$.  The occurrence of bands
with phase segregation is an indication of long range order developing
in the system as for example in thermal phase separations, though the
process here involves no thermal randomness.  In ref. \cite{molera}, a
simple Boltzmann type approximation was developed, probably the only
analytical approach available at this stage, and a linear stability
analysis could reproduce the characteristic length of the bands if $n
> .5$. The phase diagram (called the fundamental diagram) is still out
of reach of this approach \cite{molera}.

Our purpose in this paper is two fold.  One is to use this CA model
for the city traffic to explore the complexity of the jammed phases,
especially how a small change in the model can lead to drastic changes
in the steady state. Next, we identify a procedure by which an {\it
emergent} traffic jam can be predicted. This is done by analyzing the
waiting times of a tagged car while the system is evolving towards a
steady state.  Both of these features are of practical importance and
their significances are discussed at the end.
 
Our model involves an additional type of cars strictly going left.
The left cars can coexist with the right cars but can block, and can
be blocked by, an up car. See Fig. 1.  The reason for this additional
set is to introduce enough complexity in the system for the jam to
have a wider variety than a simple band.  There are, of course, more
complicated situations, but we find this to be the minimal change that
produces drastic effects.  Note that in the absence of the left cars
there is a symmetry for $\gamma <.5$ and $\gamma > .5$, and left cars
destroy this symmetry.

The evolution of a random configuration can be monitored by the
velocity that measures the average flow through a site.  The velocity
can be defined \cite{molera} as
\begin{equation}
v(t) = \frac{1}{2N} \sum_{G=R,U,L} \sum_{\bf r} 
(G_{\bf r}^{t+1} - G_{\bf r}^{t} )^2 
\end{equation}
where $N$ is the total number of cars, and $G_{\bf r}^{t}$ $=1 $ or 0
according as site ${\bf r}$ is occupied or not at time $t$ by a car of
type $G=R,U,L$, (for Right, Up or Left moving car).  A time averaged
velocity is defined as ${\bar v}=\sum_{T_1\leq t \leq T_2} v(t) /(T_2
- T_1)$.  This velocity gives a measure of the movement in the system
and attains a small value if there is any traffic jam.

We also tag one randomly chosen up car and follow its trajectory as
the system evolves.  Because of the hard core constraint, turning
probability, and signal, a car quite often stays at the same site for
a certain time interval.  Let us call this the waiting time, and let
$P(t)$ be the probability density for waiting time $t$.  We evaluate
$P(t)$ from the trajectory of the tagged car by studying the histogram
of waiting time $t$, i.e., by computing the number of times, Q, the
car stays at a site for $ t_1 < t < t_2$, so that the probability
density for a particular realization is $Q/(t_2 - t_1)$.  $P(t)$ is
then obtained by averaging over various realizations.  We show that
the $t$ dependence of $P(t)$ is different if the system evolves
towards a jammed phase than in a moving phase.  For the deterministic
model, the power spectrum for the waiting time (i.e., its Fourier
transform) at a particular site was shown to have ``$1/f$'' behavior
only at the threshold of the jamming transition\cite{nagb}.  We show
that $P(t)$ is a more powerful way of predicting an emergent jam.

In terms of the boolean variables $R, U, L$, the exact stochastic
evolution equations can be written as
\begin{mathletters}
\begin{eqnarray}
R_{\bf r}^{t+1} &=& R_{\bf r}^{t} \left [ \sigma^t {\bar \xi}_{\bf r}^t 
+  \sigma^t { \xi}_{\bf r}^t ( R_{\bf r+x}^{t} +  U_{\bf r+x}^{t})+
{\bar \sigma}^t {\xi}_{\bf r}^t + 
{\bar \sigma}^t {\bar \xi}_{\bf r}^t (R_{\bf r+y}^{t}+
U_{\bf r+y}^{t}) \right ]
+ \nonumber \\
&&{\bar R}_{\bf r}^t {\bar U}_{\bf r}^{t} \big [
\sigma^t { \xi}_{\bf r-x}^t  R_{\bf r-x}^{t} +
{\bar \sigma}^t {\bar \xi}_{\bf r-y}^t  R_{\bf r-y}^{t} \big ] \\
U_{\bf r}^{t+1} &=& U_{\bf r}^{t} \big [ {\bar \sigma}^t 
{\bar \eta}_{\bf r}^t 
+ {\bar \sigma}^t { \eta}_{\bf r}^t W_{{\bf r+y}}^t +
{ \sigma}^t {\eta}_{\bf r}^t + 
{ \sigma}^t {\bar \eta}_{\bf r}^t W_{{\bf {r+x}}}^{t} \big ]
+\nonumber\\
&& {\bar R}_{\bf r}^t {\bar U}_{\bf r}^{t}{\bar L}_{\bf r}^{t} \left [
{\bar \sigma}^t { \eta}_{\bf r-y}^t  U_{\bf r-y}^{t} +
{\sigma}^t {\bar \eta}_{\bf r-x}^t  U_{\bf r-x}^{t} \right ],\\
L_{\bf r}^{t+1} &=& L_{\bf r}^{t} \sigma^t ( L_{\bf r-x}^{t} +  
U_{\bf r-x}^{t} - L_{\bf r-x}^{t} U_{\bf r-x}^{t})
+ {\sigma}^t {\bar U}_{\bf r}^{t}{\bar L}_{\bf r}^{t} L_{\bf r+x}^{t}+
{\bar \sigma}^t L_{\bf r}^{t},
\label{eq:bool}
\end{eqnarray}
\end{mathletters}
where $\sigma^t = t\ {\rm mod} 2$ is the signal, $\xi_{\bf r}^t = 1 \
{\rm or} 0$ with probability $1-\gamma$ or $\gamma$ denotes turning of
a right type car at site ${\bf r}$ and time t, $\eta$ as the
corresponding variable for the up cars, ${\bf x,y}$ are the unit
vectors for nearest neighbors, ${\bar a} = 1 - a$ for any $a$, and
$W_{{\bf r}}^t = U_{\bf r}^{t} + L_{\bf r}^{t} +R_{\bf r}^{t} - R_{\bf
r}^{t} L_{\bf r}^{t} - U_{\bf r}^{t} L_{\bf r}^{t} $ is a combination
boolean variable for site $\bf r$.  These equations can be obtained by
considering the various possibilities of a car staying at the same
site or coming from a neighboring site.  The configuration at time
$t+1$ is determined by the configuration at time $t$, i.e., each car
moves according to the position of the cars at time $t$.

A hard core constraint requires that $R_{\bf r}^t U_{\bf r}^t = 0$.
We also have $U_r^t L_r^t =0$ except for a $U$-hole-$L$ type
configuration during a horizontal move (Fig. 1). The boolean variable
$W$ takes care of this eventuality. $W$ is 1 if the site is occupied
by any type of car and 0 if it is unoccupied.  We have numerically
evolved the system from arbitrary initial configurations for lattices
upto $64\times 64$ with periodic boundary conditions.  We first
present the results and then discuss them.

{\em{Deadlock Phase :}} With no left cars, it is known that there is a
critical density above which traffic jam occurs in the form of bands
\cite{molera}.  We chose a density ($n=.67$) higher than the critical
value.  As the left car density is increased we observe a roughening
of the edges of the bands.  In fact, if the band width is small so
that there are two bands in the lattice \cite{comm2}, then beyond a
certain left car density, the band changes to a single one indicating
that the bandwidth increases as the density increases.  A more drastic
change takes place at a still higher density (though much smaller
compared to the overall density) when the structure loses the band
pattern.  This is a new phase that we call a {\it deadlock phase}.
The structure is bounded by vertical lines on the left side and left
cars on the right, as shown in Fig. 2.  The deadlock phase has a
strictly zero velocity compared to the band phase where there is a
residual velocity $(\sim N^{-1/2})$ coming from the edges of the
bands.

The evolution has been repeated several times and we find a broad
region of left densities where the steady state can be of either type.
Such coexistence is a reflection of a first order transition. In fact
for $\gamma =.8$, there is a range of left car densities where we have
observed sequentially the coexistence of ``two band''- ``one band''
phases, two band - one band - deadlock phases, and one band-deadlock
phase before going over to the deadlock phase.

For $\rho_l =0$, there is a symmetry between $\gamma < .5$ and $\gamma
> .5$, but not for $\rho_l \neq 0$ because of the hardcore repulsion
between U and L.  For $\gamma >.5$, the segregation of U and R is
opposite to that for $\gamma <.5$.  For a small number of left cars,
with $\gamma > .5$, the L's settle down at the RU interface, and as the
number increases they expose some of the U's.  This leads to the
roughening of the edge leading to the deadlock phase.  See Fig 2.
The density of left cars for the transition from band to deadlock 
is a monotonically increasing function of $\gamma$.  We do not go into
the details of  these results  because of the large coexistence region.

{\it {Waiting time:}} We now tag one up car at random from the
beginning of the evolution.  The trajectory of the tagged particle
under various conditions are shown in Fig 3. We observe that in the
moving phase the car traverses a major part of the city while the
motion is restricted to bands in Figs 3b-e.  There is no such definite
pattern for the deadlock phase as shown in Figs 3e,f. From these
trajectories we computed the waiting probability, P(t).  When there is
no jam, the tagged car does not spend much time at the crossings and
$P(t)$ decays fast, almost exponentially. This is shown in Fig. 4a for
$\gamma =.5$ without any left cars.  For values of $\gamma$, when
there is a band phase or a deadlock phase, the waiting time seems to
have a power law decay $P(t) \sim t^{-w}$ with $w=2.4$ for both the
jammed states. See Fig 4.  It therefore appears that, for an emerging
jammed situation, the waiting time for a car has no definite time
scale. Haven't we all have that feeling when stuck in a jam?

Let us now discuss these results. First, we have shown that the simple
phase diagram can be modified drastically by the addition of a {\it
small} density of left cars.  Our simulations are at a density for
which, except for $\gamma = .5$, bands are supposed to form along the
diagonal in absence of any left cars.  In the deadlock phase, there is
a vertical wall on the left side but the left cars bunch together on
the right side.  Since the left cars, by construction, are restricted
to one dimension, we concentrate on the motion of the left cars of one
row only. These cars can hop to the left, only if the site is not
occupied by a U or L car.  Note that the left cars are transparent to
the right cars.  If, in the spirit of the Boltzmann approximation, the
effect of the up cars is taken into account by a random blocking of a
vacant site with probability $n/2$ (the density of up cars), {\it
uncorrelated} both in time and space, no bunching can occur (we have
also checked it explicitly).  In other words, the bunching of the left
cars we observe is mainly due to a correlated blocking both in space
and time.  More quantitatively, one might use the Boltzmann
approximation of Ref. \cite{molera} to study the stability of a
homogeneous phase.  In this approximation, the exact stochastic
equations are replaced by time evolution equations for the average
densities, ignoring all correlations.  From Eq. 2, one sees that a
small density of left cars would be a small perturbation to the
equations derived in Ref. \cite{molera}. Such a small perturbation
does not destroy the instability towards a band phase, and, hence,
cannot yield a deadlock phase.

Similarly, one can invoke a Boltzmann approximation for $P(t)$ also.
An up car at time $t+1$ will stay at its site if {\it (i)} the sites
it can move to are blocked, {\it (ii)}signal stops it, or {\it (iii)}
the turning probability does not allow it to move.  Therefore, the
probability of staying over at the same site at time $t$ is
proportional to
\begin{equation}
p({\bf r},t) \equiv \big [ {\bar \sigma}^t 
{\bar \eta}_{\bf r}^t 
+ {\bar \sigma}^t { \eta}_{\bf r}^t W_{{\bf r+y}}^t +
{ \sigma}^t {\eta}_{\bf r}^t + 
{ \sigma}^t {\bar \eta}_{\bf r}^t W_{{\bf {r+x}}}^{t} \big ].
\end{equation} 
It then follows that the probability of waiting time $t$ at a site
${\bf r}$, if the up car comes to the site at time $T$, is
proportional to $\prod_{m=0}^{t} p({\bf r},T+m) $.  $P(t)$ is now
obtained by averaging over all sites, all $T$ and all realizations.
In the simplest situation if we replace all the boolean variables by
the averages as in the Boltzmann approximation \cite{molera}, we see
that in a moving phase ( in absence of any left cars) $P(t) \sim
[(1+n)/2]^t \sim \exp (-t \mid\ln (1+n)/2\mid)$.  This shows an
exponential tail consistent with our observation for the moving phase.
Inclusion of a small density of left cars do not change this result
significantly.  To get a power law tail, again the correlations have
to play a role.

For one dimensional highway traffic, the life time of jammed regions
has been found to have a power law tail and this has been attributed
to the avalanches or self organized critical (SOC) behavior of the
model\cite{nagelb}.  This SOC behavior is characteristic of the steady
state of the 1-d model. In the two dimensional case, we are observing
the power law behavior in the {\it approach} to the steady state.  It
is tempting to associate an underlying SOC type behaviour as the
origin of this power law for waiting times, but such an identification
is not apparent.
 
The bunching of the left cars on the right edge in the deadlock phase
has another implication.  If these left cars are removed from the
deadlock structure, the jammed phase goes into {\it one} single band
structure along the diagonal, not necessarily the steady state
solution of the original zero left car density case.  The deadlock
phase is a compact lattice spanning structure, and to create more than
one band, it is necessary to produce cracks (i.e. vacant sites) in the
middle of the structure, which due to hard core repulsion and tight
packing (see Fig 2) is impossible.  An interesting way to look at this
is to think of the minute quantity of left cars as acting as an
adhesive to bring together the bands.  Since, the left cars get
segregated in any case, their removal from the road is rather easy,
and the system relaxes to a single structure. This might be a
practical way of bringing together the jammed region into the central
part of the city relieving all other regions.  We wonder if this helps
in real life in any way.

To summarize, we have shown that the band phase for two types of cars
can be modified drastically by addition of small number of left moving
cars.  The bunching of the left cars in the deadlock phase is
significant.  We have also shown that the waiting time has a power law
distribution as the traffic jam is approached. In contrast, in the
moving phase, the long waiting times are almost not present. This
gives a useful way of {\it predicting beforehand } if a traffic jam is
emerging or not. An experimental verification of this prediction would
be highly interesting.

\psfig{file=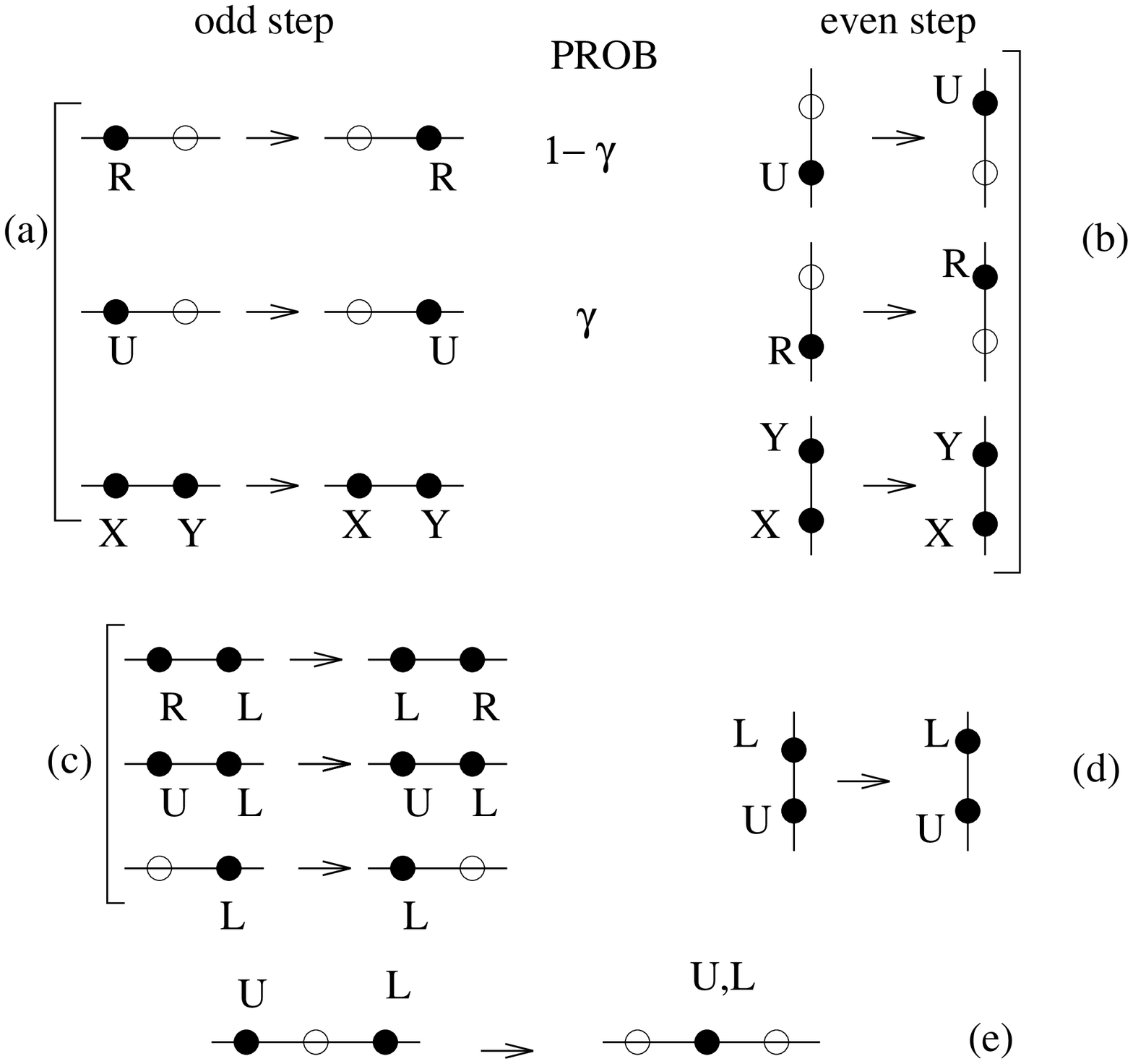}
\begin{figure}
\caption{The possible moves for odd steps, and even steps. (a) and (b) 
correspond to the case with no left cars.  The new moves with left cars
are (c), (d) and (e). (e) is special because it should not have occured 
if strict hard core constraint is used. Since the cars do not look beyond 
nearest neighbors, this has to be accepted. X, Y stand for any type of car.
}
\end{figure}
\newpage
\psfig{file=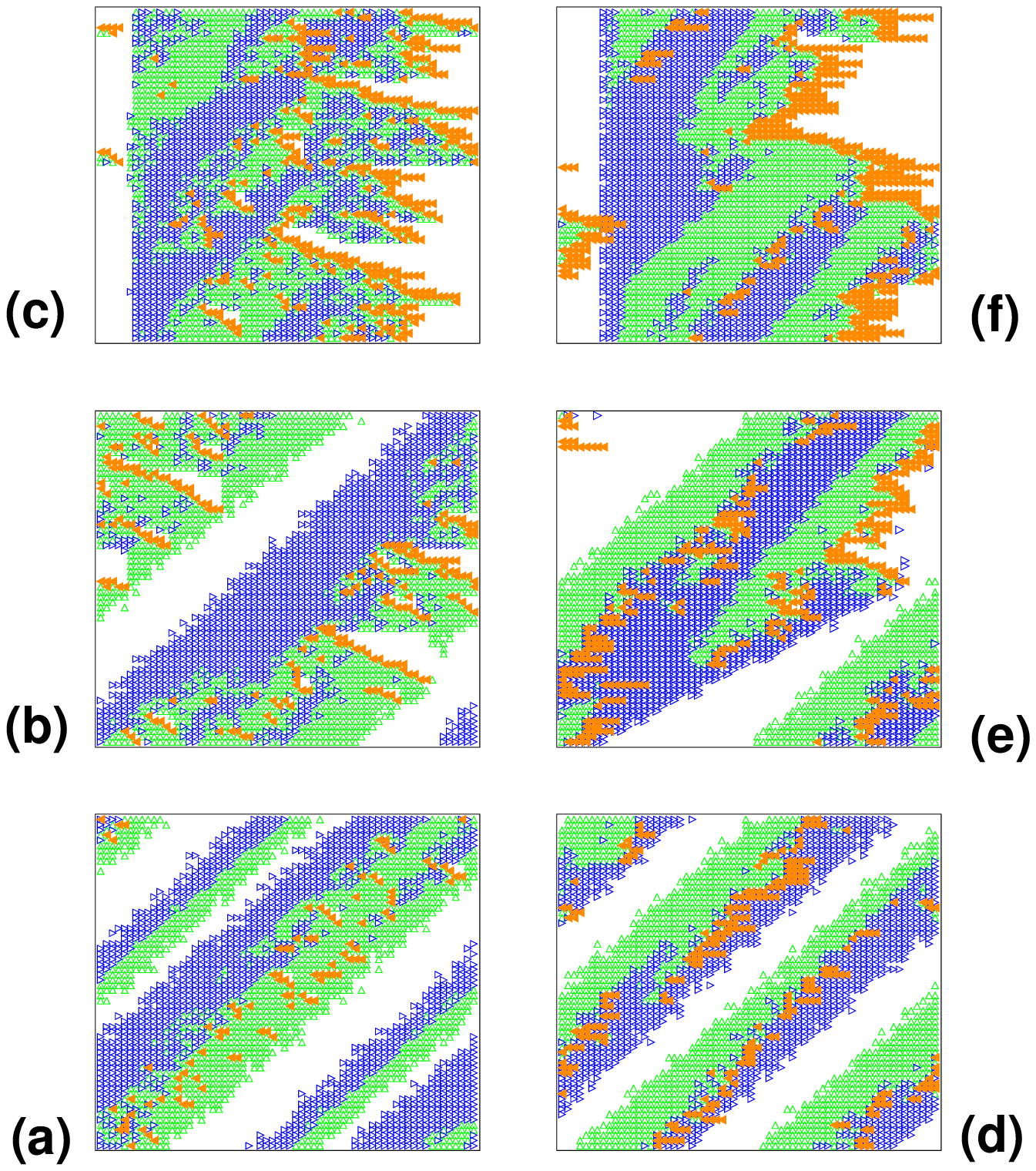}
\begin{figure}
\caption{ The steady state phase changes from two band (a,d) to
one band (b,e) and to deadlock (c,f) as the number (\#) of left cars 
(red) increases.  All are for $64\times 64$ lattices
with periodic boundary conditions, and 2730 right (blue) and up (green 
) cars. 
(a, b, c) $\gamma=.2$, $\#= 100,220,420$. (d, e, f) $\gamma=.8$, $\#=270,
400, 500$.  Note the reversal 
in the car pattern for $\gamma >.5$. Also note the vertical wall on the left 
in (c) and (f).}
\end{figure}
\newpage
\psfig{file=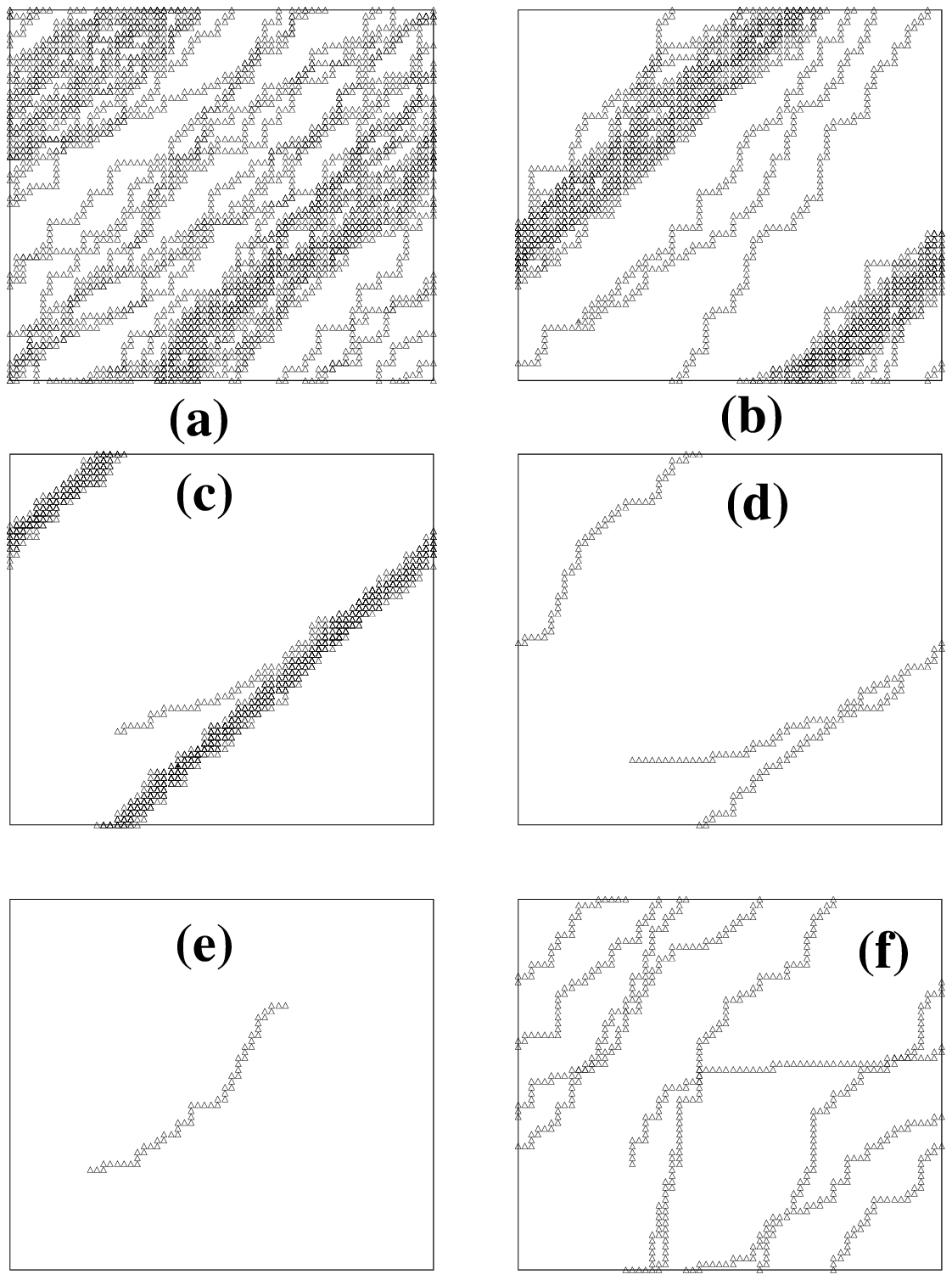}
\begin{figure}
\caption{The path of a tagged ``up'' car, for several values of
$\gamma$ and the number (\#)of left cars.
(a) When there is no jam.  This actually
corresponds to $\gamma=.5$ with no left cars. (b) When the jam consists of 
one single band. $\gamma=.6, \ \#=160$. 
(c,d) When the jam consists of two bands for $\gamma=.8, \ \# =0$. (c) shows
a car getting into the middle band, while (d) shows, from a different
realization, a car getting into the outer band.
(e,f) Now the jammed state is a deadlock phase, for $\gamma=.4, \ \#=270 $.
Two possible 
paths are shown from two different samples. In case (e) the total path is 
rather small compared to (f).  The starting point is always chosen,
for convenience, near the lower left part of the lattice.}
\end{figure}
\newpage
\psfig{file=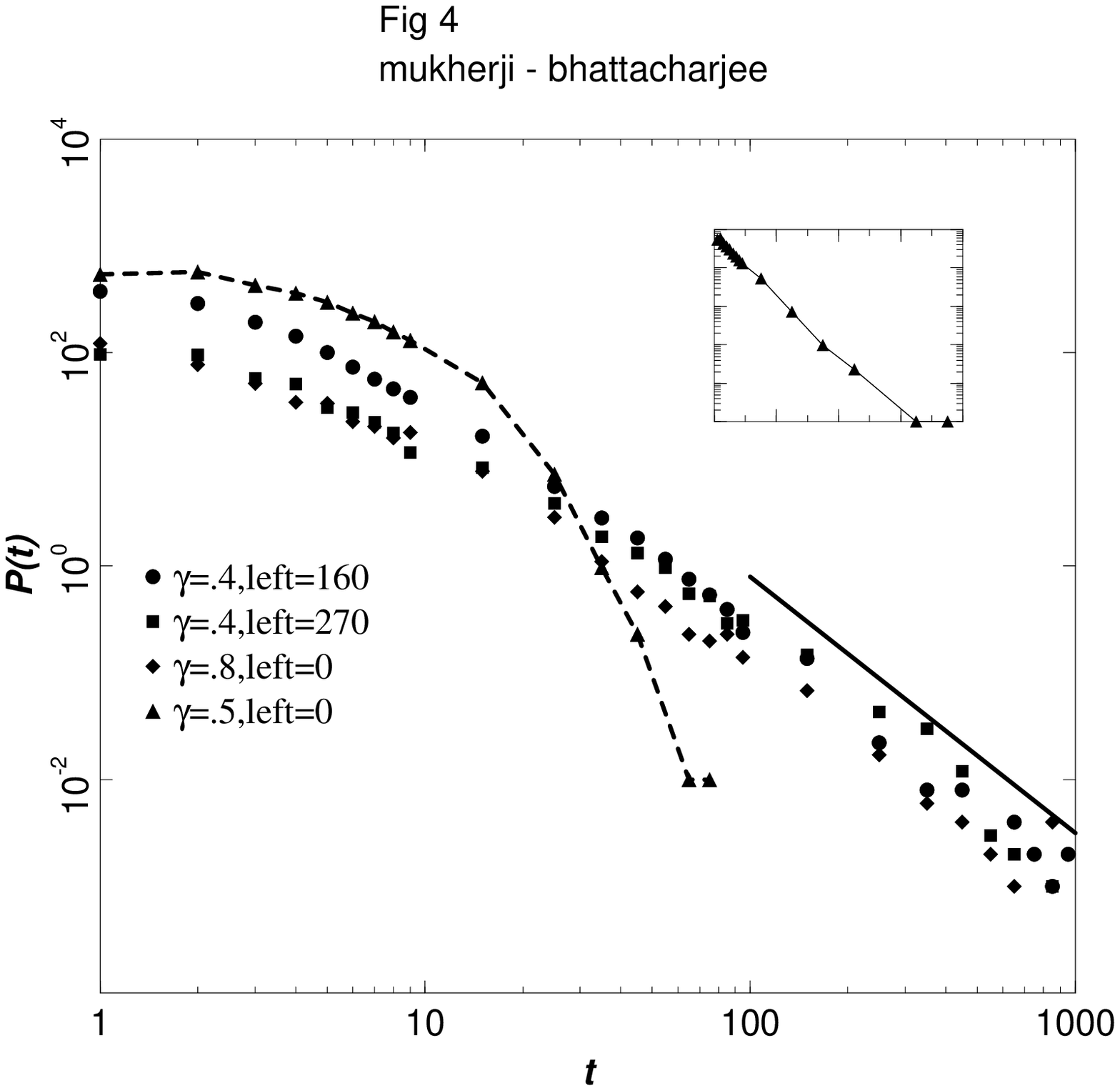}
\begin{figure}
\caption{P(t) vs. t Plot.  The values of $\gamma$ and the number of left cars
are given in the legend.  
The solid line has a slope of 2.4, indicating $P(t) \sim t^{-2.4}$. 
The plot for the moving phase represented by the 
dashed line is shown in the inset on a semilog scale. }
\end{figure}


\begin{references}
\bibitem{comm}This particular traffic model has similarities with the
two species driven lattice gas of Ref. \cite{ziats} with $\gamma$
playing the role of the Boltzmann factor.
\bibitem{comm1} It should be added that the cases $\gamma=0$
(deterministic) and $\gamma \rightarrow 0$ are not identical.  See Ref
\cite{martinez}.
\bibitem{comm2}If the bandwidth is $w$, then for an $N\times N$ lattice, the 
number of bands one expects to see is the integer part of $N/w$.
\end{references}
\end{document}